\documentclass[prd,twocolumn,tightenlines,superscriptaddress,longbibliography]{revtex4-1}

\usepackage[utf8]{inputenc}

\usepackage[T1]{fontenc}
\usepackage{bbold}
\usepackage{amssymb}
\usepackage{amsmath}
\usepackage{graphicx}
\usepackage{color}
\usepackage[colorlinks=true,linkcolor=blue,citecolor=red,urlcolor=blue]{hyperref}
\usepackage[section]{placeins}

\usepackage{enumitem,amssymb}
\newlist{todolist}{itemize}{2}
\setlist[todolist]{label=$\square$}
\usepackage{pifont}

\newcommand{\be}{\begin{equation}}
	\newcommand{\ee}{\end{equation}}

\newcommand{\bra}[1]{\langle{#1}|}
\newcommand{\ket}[1]{|{#1}\rangle}

\usepackage{ulem}

\begin{document}
	
	\title{Topological edge states in a Rydberg composite}
	
	\date{\today}

	\author{Matthew T. Eiles}
	\email{meiles@pks.mpg.de}
	\affiliation{Max-Planck-Institut f\"ur Physik komplexer Systeme, N\"othnitzer Str.\ 38, 
		D-01187 Dresden, Germany }

	\author{Christopher W. W\"achtler}
	\affiliation{Max-Planck-Institut f\"ur Physik komplexer Systeme, N\"othnitzer Str.\ 38, 
		D-01187 Dresden, Germany }
	\affiliation{Department of Physics, University of California, Berkeley, California 94720, USA }
	
	\author{Alexander Eisfeld}
	\affiliation{Max-Planck-Institut f\"ur Physik komplexer Systeme, N\"othnitzer Str.\ 38,
		D-01187 Dresden, Germany }
	\affiliation{Universität Potsdam, Institut für Physik und Astronomie, Karl-Liebknecht-Str. 24-25, 14476 Potsdam, Germany}
	
	\author{Jan M. Rost}
	
	\affiliation{Max-Planck-Institut f\"ur Physik komplexer Systeme, N\"othnitzer Str.\ 38,
		D-01187 Dresden, Germany }

	\begin{abstract}
		We examine topological phases and symmetry-protected electronic edge states in the context of a Rydberg composite: a Rydberg atom interfaced with a structured arrangement of ground-state atoms.
		The electronic Hamiltonian of such a composite possesses a direct mapping to a tight-binding Hamiltonian, which enables the realization and study of a variety of systems with non-trivial topology by tuning the arrangement of ground-state atoms and the excitation of the Rydberg atom.
		The Rydberg electron moves in a combined potential including the long-ranged Coulomb interaction with the Rydberg core and short-ranged interactions with each neutral atom; the effective interactions between sites are determined by this combination. 
		We first confirm the existence of topologically-protected edge states in a Rydberg composite by mapping it to the paradigmatic Su-Schrieffer-Heeger dimer model. 
		Following that, we study more complicated systems with trimer unit cells which can be easily simulated with a Rydberg composite. 
	\end{abstract}
	\maketitle
	Topological insulators \cite{mooreNext2009,mooreBirth2010,hasanThreeDimensional2011} describe a special class of solids exhibiting an insulating bulk but conducting surface states. 
	These display a surprising immunity to a wide range of local deformations, inherently avoiding backscattering over broad energy ranges and circumventing localization in the presence of disorder.
	Questions about the existence, behavior, and characterization of topological insulators and the symmetry-protected edge states that they can host have motivated rapid growth in this field in recent years \cite{mittalTopologically2014,lohseThouless2016, nakajimaTopological2016, lohseExploring2018,krausTopological2012,verbinTopological2015,zilberbergpho2018,kaneTopological2014,pauloseTopological2015,salernoDynamical2014,susstrunkObservation2015,ningyuanTime2015,albertTopological2015,aspuru-guzikPhotonic2012,kitagawaObservation2012,zhaoTopological2018,arkinstallTopological2017,zhangExperimental2019,yuen-zhouTopologically2014}.
	A major effort in this direction is the exploration of well-controlled systems that exhibit novel topological properties and can be used to clarify questions about the behavior of topological invariants. 
	Ultracold Rydberg atoms are promising quantum simulators in this respect due to their high controllability and exaggerated properties. 
	Two recent examples using very different approaches illustrate this in the context of the {Su-Schriefer-Heeger (SSH)} model \cite{suSolitons1979a}.
	In the first, several Rydberg atoms were arranged in an optical tweezer array. 
	The long-range dipolar interactions between two Rydberg states of different angular momentum enabled the desired staggered hopping amplitudes \cite{deleseleucObservation2019a}.
	In the second example, multiple Rydberg levels of a single atom were employed to form a synthetic one-dimensional lattice, with microwave coupling between these levels setting the hopping amplitudes \cite{kanungoRealizing2022b}.
	The design of simulators of topologically richer systems than the SSH model remains an area of active research, with most proposals utilizing the long-range interactions between Rydberg atoms prepared in complicated geometries \cite{yangQuantum2022,samajdarQuantum2021a,liSymmetryProtected2021a,verresenUnifying2022,giudiciDynamical2022a,samajdarEmergent2023,weberExperimentally2022a}.

	In this paper, we introduce a different approach utilizing a \textit{single} Rydberg atom and an ensemble of trapped ground-state atoms (scatterers) located within the Rydberg electron's orbit. 
	Confinement of these scatterers in a particular arrangement can be provided by an optical lattice or in an array of optical tweezers. 
	Such a \textit{Rydberg composite} allows us to design Hamiltonians which exhibit features associated with symmetry-protected topological insulators.
	As we will show, 
	the topological properties can be tuned by the choice of the scatterer positions and by the principal quantum number $\nu$ of the Rydberg atom. 
	To demonstrate how to use a Rydberg composite to study topological physics, we show how to realize three different lattices of increasing complexity.

	\begin{figure*}[t]
		\includegraphics[width=0.95\textwidth]{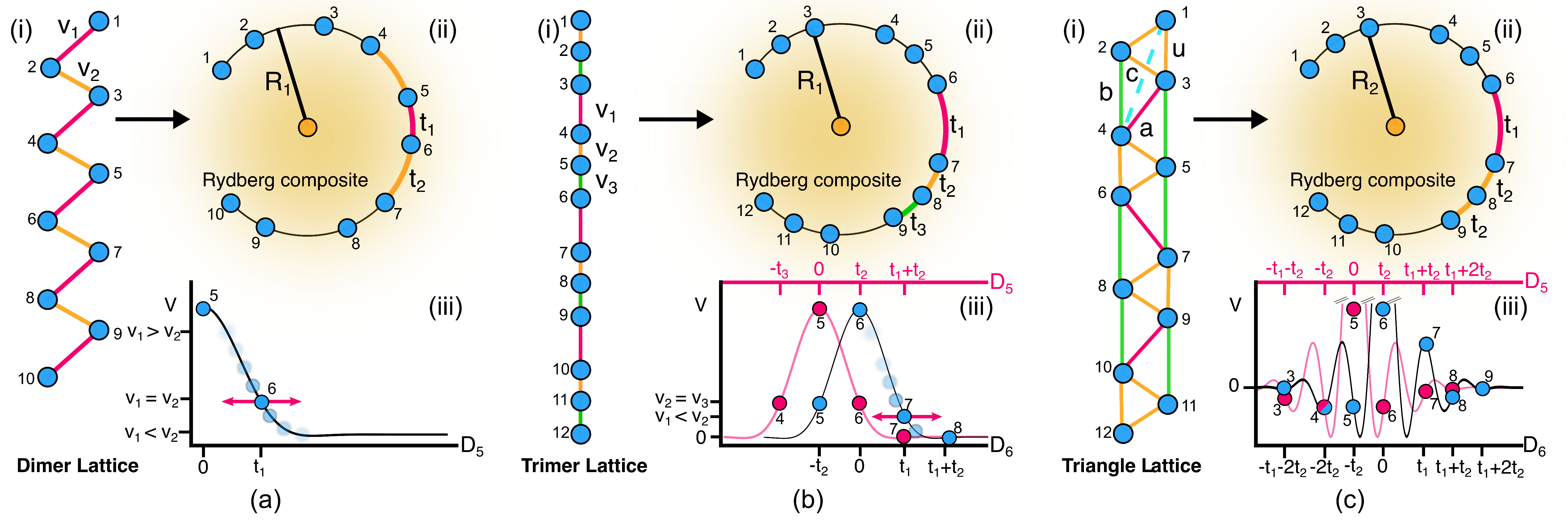}
		\caption{\label{fig:intro}
			Schematics of the three models: (a)  dimer SSH chain, (b) trimer SSH chain, and (c) triangle chain. In each panel, (i) depicts the desired tight-binding lattice and defines the couplings between sites, and (ii) sketches the Rydberg composite  corresponding to this lattice. The arc lengths shown in this panel determine the effective couplings, as illustrated in each panel by (iii), which shows the relevant interaction curves.  The interaction curve $V$ in (a)iii shows the hopping amplitude between site $5$ and neighboring sites as a function of $D_5$, the distance around the circle away from site $5$. As the position of site 6 moves, so does the energy $V_{56}$ (blue circle marked 6), modifying the ratio $v_2/v_1$. Two different interaction curves are shown in (b)iii. The black (pink) curve shows the interaction $V$ as a function of the distance $D_6$ ($D_5$) away from site 6 (5).  The blue markers denote the on-site potential $E_6$ and hopping amplitudes $V_{65}$, $V_{67}$, and $V_{68}$; the pink markers show $E_5$, $V_{54}$, $V_{56}$, and $V_{57}$.	The next-nearest-neighbor amplitudes $V_{68}$ and $V_{57}$ are negligible. By varying $t_1$, the distance between sites $6$ and $7$ is changed, which modifies $V_{67}$ while keeping constant the other interactions.  Panel (c)iii shows these same interaction curves, but now in the more complicated geometry relevant to the triangle lattice. Careful inspection of the different points on these two curves shows how the hopping amplitudes $u$, $a$, $b$, and $c$ in the triangle lattice are realized; the amplitude $u$ is fixed for this geometry, while the other amplitudes vary as a function of $t_1$. 
		}. 
	\end{figure*}
	
	The effective Hamiltonian \cite{eilesUltracold2016,anderson,hunterRydberg2020a,eilesRing2020} of a Rydberg composite, with scatterers placed at positions $\vec{R}_q$, is
	\be
	\label{eq:rydham}
	H_\mathrm{e} = -\frac{\nabla^2}{2} -\frac{1}{r} + \sum_{q = 1}^M2\pi a_s\delta^3(\vec r - \vec R_q)
	\ee
	in atomic units. 
	The first two terms govern the electron's motion in the Coulomb field of the Rydberg core, while the last term describes its interaction with the scatterers using the Fermi pseudopotential, valid in the low-energy scattering limit.
	The interaction strength is determined by the $S$-wave scattering length $a_s$ \cite{greeneCreation2000a,eilesTrilobites2019}. 
	We neglect the role of quantum defects, as we focus on the perturbation of the degenerate hydrogen-like states with angular momentum greater than $3$ \cite{anderson}. 
	To facilitate the choice of scatterer positions to realize a particular topologically interesting system, it is convenient to transform the Hamiltonian (\ref{eq:rydham}) to the form of a tight-binding Hamiltonian.
	This mapping relies on the fact that
	the electron-scatterer interaction \cite{fermiSopra2008,omontTheory1977} is too weak to mix Rydberg states with different principal quantum numbers $\nu$, but nevertheless it splits the degenerate energy levels with different angular momentum $l$ but the same $\nu$ into two subspaces. 
	One, of size $\nu^2-M$, remains degenerate and unshifted, while the second, of size $M$, splits away \cite{anderson,eilesTrilobites2019}.
	The spectrum of the Hamiltonian $H_\mathrm{e}$ restricted to this non-trivial subspace coincides exactly with that of the Hamiltonian
	\begin{equation}
		\label{eq:tightbindinghamiltonian}
		H = \sum_{q}^ME_q\ket{q}\bra{q} + \sum_q^M\sum_{q'\ne q}^MV_{qq'}\ket{q}\bra{q'}.
	\end{equation}
	The states $\ket{q}$ describes a wave function which is localized on the scatterer at position $\vec{R}_q$.
	The matrix elements $E_q$ and $V_{qq'}$ are determined by the positions of the scatterers and the principal quantum number $\nu$	
	\footnote{
		In the high $\nu$ limit where the effect of quantum-defect-shifted states is negligible, the matrix elements of Eq.\ref{eq:tightbindinghamiltonian} are given in closed form: 
		\begin{align} 
			\label{eq:energydefs} 
			V_{qq'}&=  \frac{u_{\nu 0}'(x_-)u_{\nu 0}(x_+) - u_{\nu 0}(t_-)u_{\nu 0}'(x_+)}{2(x_+-x_-)}.
		\end{align}
		Here, 
		$x_\pm = \frac{1}{2}\left(R_q +R_{q'} \pm |\vec R_q - \vec R_{q'}|\right)$, $R_q = |\vec R_q|$, and $u_{\nu 0}(r)$ is the $s$-wave reduced hydrogen radial function defined in the usual way, 
		$\langle\vec{r}|\nu l m\rangle = \frac{1}{r}u_{\nu l}(r) Y_{lm}(\hat r)$.  
		The prime denotes the spacial derivative.}.
	Further details about the connection of this Hamiltonian with Eq.~(\ref{eq:rydham}) can be found in Refs.~\cite{hunterRydberg2020a, anderson, eilesRing2020}.

	The three composites illustrated in Figure~\ref{fig:intro} exemplify the ability of different scatterer arrangements, in combination with the choice of Rydberg state $\nu$, to design and realize different effective lattice Hamiltonians. 
	We will use these three composites to demonstrate topological physics in a Rydberg composite. 
	In each case, we consider atoms arranged in a (broken) ring around the Rydberg atom, which fixes a common $E_q$ for all scatterers.
	We select the number of scatterers $M$, the principal quantum number $\nu$, and radius $R=R_q$ such that the desired hopping terms $V_{qq'}$ are realized.
	To confirm that this leads to a topologically non-trivial configurations, we theoretically analyze each setup by applying periodic boundary conditions to the effective Hamiltonian, giving insight into its {topological} bulk properties. 
	The bulk-boundary correspondence allows us to subsequently ascertain its topological aspects in the finite system.

\begin{figure}[t]
		\includegraphics[width=0.95\columnwidth]{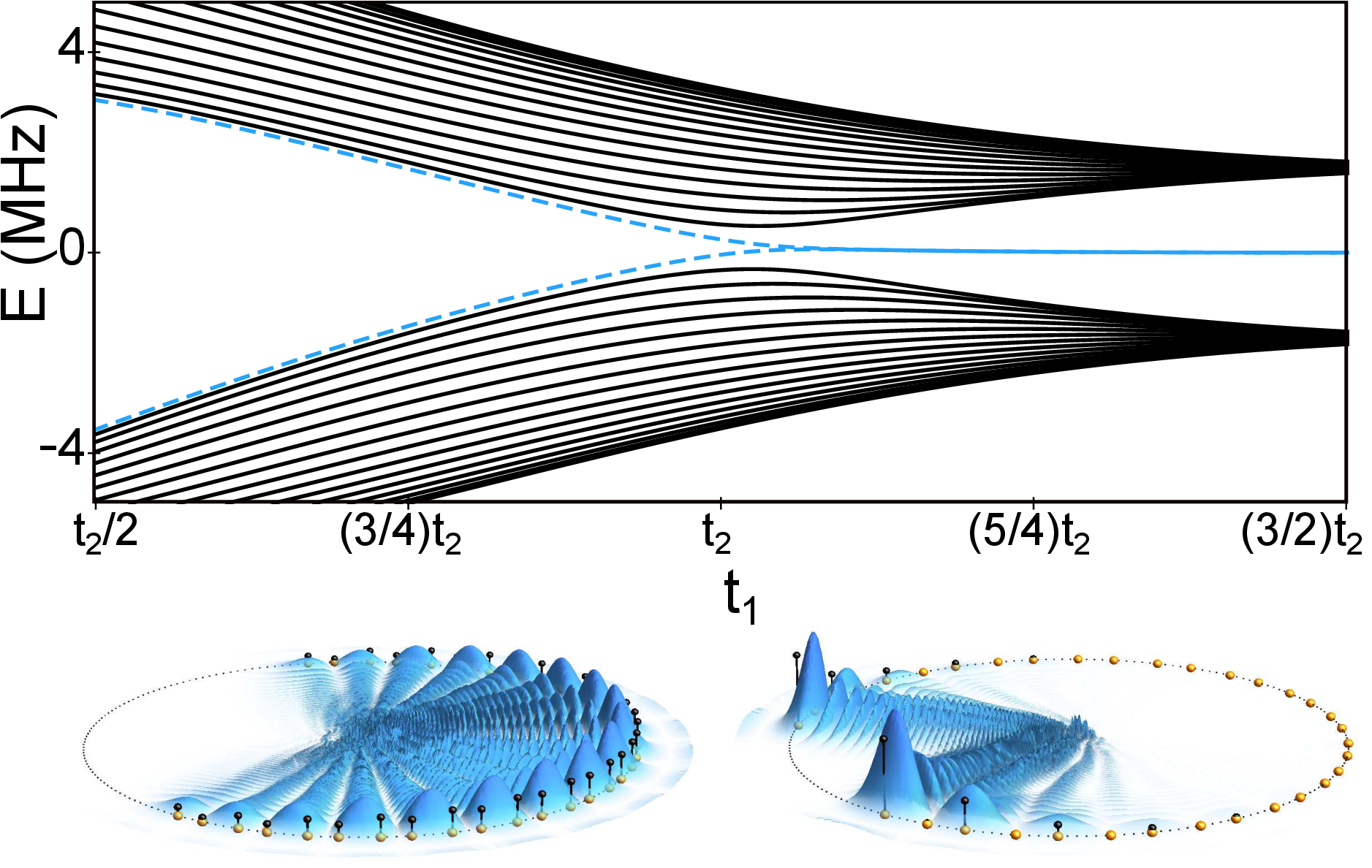}
		\caption{\label{fig:SSH}
			The energy spectrum of the SSH Rydberg composite, where $\nu = 60$ and $R = 2\nu^2$. The energies are centered around the on-site potential $E_q$, and are plotted as a function of $t_1$ for fixed $t_2= 2\pi R/45$. 
			The exemplary wave function images shown below the spectrum illustrate the bulk eigenstate corresponding to the dashed-blue level in the upper band (left,  $t_1 = 3t_2/4$) and an edge state (right, $t_1 = 5t_2/4$).
		}
	\end{figure}

	The first model that we consider is the dimer SSH model depicted in Fig.~\ref{fig:intro}a(i): a one-dimensional lattice with staggered nearest-neighbor hopping amplitudes  $v_1$ and $v_2$ \cite{suSolitons1979a}. 
	This paradigmatic model introduces many concepts useful in the analysis of more complicated systems. 
	Fig.~1a(ii) shows how to design a Rydberg composite which realizes this model.
	We set $R = R_1 = 2\nu^2$ and place the scatterers on this ring so that they are separated by arclengths $t_1$ and $t_2$. 
	By choosing $M<\nu$ we guarantee that the scatterers are spaced sufficiently far apart that $V_{qq'}$ is negligible when $|q'-q|>1$ \cite{anderson}. 
	Then, the monotonic dependence of $V_{qq'}$ on $t_1$ allows us to stagger the hopping elements simply by staggering the distances $t_1$ and $t_2$, as is illustrated in Fig.~\ref{fig:intro}(a)iii. 
	For illustrative purposes we choose $\nu = 60$, $M = 36$, and fix $t_2 = 2\pi R/45$.  
	These choices yield a reasonably large lattice size without making the resulting band spectrum and wave functions overly complicated to visualize. 
	Fig.~\ref{fig:SSH} shows the resulting eigenspectrum as a function of $t_1$. 
	It consists of two distinct bands separated by an empty band gap when $t_2>t_1$, i.e. when $v_1>v_2$ (see Fig.~\ref{fig:intro}a(i)). 
	This band gap closes when $t_1 = t_2$. 
	When $t_1>t_2$, $v_2>v_1$ and the two levels closest to the band gap, highlighted in blue, split from the bulk and become degenerate at the center of the spectrum. 
	Investigation of the eigenstates of these levels shows that they evolve from bulk states spread over the entire composite when $t_2<t_1$ to edge states appearing at the boundary of the chain of scatterers when $t_1>t_2$. 
	In the two eigenstates depicted in Fig. \ref{fig:SSH}, the black spheres represent the wave function in the site basis used to describe $H$ in Eq. (\ref{eq:tightbindinghamiltonian}), while the blue wave function shows the full electronic wave function in position space
	\footnote {We plot the absolute values of the wave functions here.}.

	The existence of this transition from bulk states living in the energy bands to edge states situated at the center of the band gap can be understood  after considering the bulk momentum Hamiltonian 
	\be
	\label{eq:ssh}
	H_\text{SSH}(k) = \begin{pmatrix}
		0 &  v_1+v_2e^{-ik}\\ 
		v_1+v_2e^{ik} & 0
	\end{pmatrix},
	\ee
	which is obtained by applying periodic boundary conditions to the lattice in Fig.~\ref{fig:intro}(a)i. 
	For the one-dimensional systems studied here, the Zak phase, 
	\be
	\label{eq:Zak}
	\mathcal{Z} = i\int_{-\pi}^\pi\langle \psi_k|\partial_k\psi_k\rangle dk,
	\ee
	where $\ket{\psi_k}$  is an eigenstate of $H(k)$,
	is a quantized topological invariant which can only take the values zero or $\pi$ (modulo $2\pi$) as long as a symmetry is present \cite{zakBerry1989}.
	The chiral symmetry of the SSH model ensures that such a topological invariant exists. 
	Computation of the Zak phase using the eigenstates of Eq.~\ref{eq:ssh} gives $\mathcal{Z} = \pi$ when $t_2>t_1$ and $\mathcal{Z} = 0$ when $t_2 < t_1$ \cite{restaMacroscopic1994,restaManifestations2000} \footnote{
		We evaluate $\mathcal{Z}$ on a discrete mesh of $k$ points using
		\be
		\mathcal{Z} = -\text{Im}\ln\Pi_k\langle \psi_k|\psi_{k+1}\rangle,
		\ee
		which does not require the phase of the eigenstates $|\psi_k\rangle$ to be a continuous function of $k$. 
	}.

	The Zak phase is intimately related to the existence of edge states through the bulk-boundary correspondence: the change in the Zak phase at $t_1 = t_2$ reveals a topological phase transition and the appearance of edge states in the finite-sized system  \cite{suSolitons1979a,zakBerry1989}.
	The implications of these edge states are manifold.
	If the scatterers are not perfectly positioned at their angles on the ring, the hopping amplitudes $V_{qq'}$ become disordered but the chiral symmetry of the lattice is preserved. 
	In such a case, although states in the bands will become localized due to this disorder, these edge states will remain unperturbed due to the symmetry protection afforded by the topology of the system. 
	In contrast, disorder of the atom positions in the radial direction leads to diagonal disorder and a breakdown of the chiral symmetry. 
	Consequently, the edge states are no longer symmetry protected and become indistinguishable from the band. 
	Moving beyond the SSH model or random disorder, one could  modify the on-site potentials in a controlled fashion by shifting the scatterers to slightly different ring radii, therefore simulating the Rice-Mele model \cite{RiceMele}.

\begin{figure}[t]
		\includegraphics[width=0.95\columnwidth]{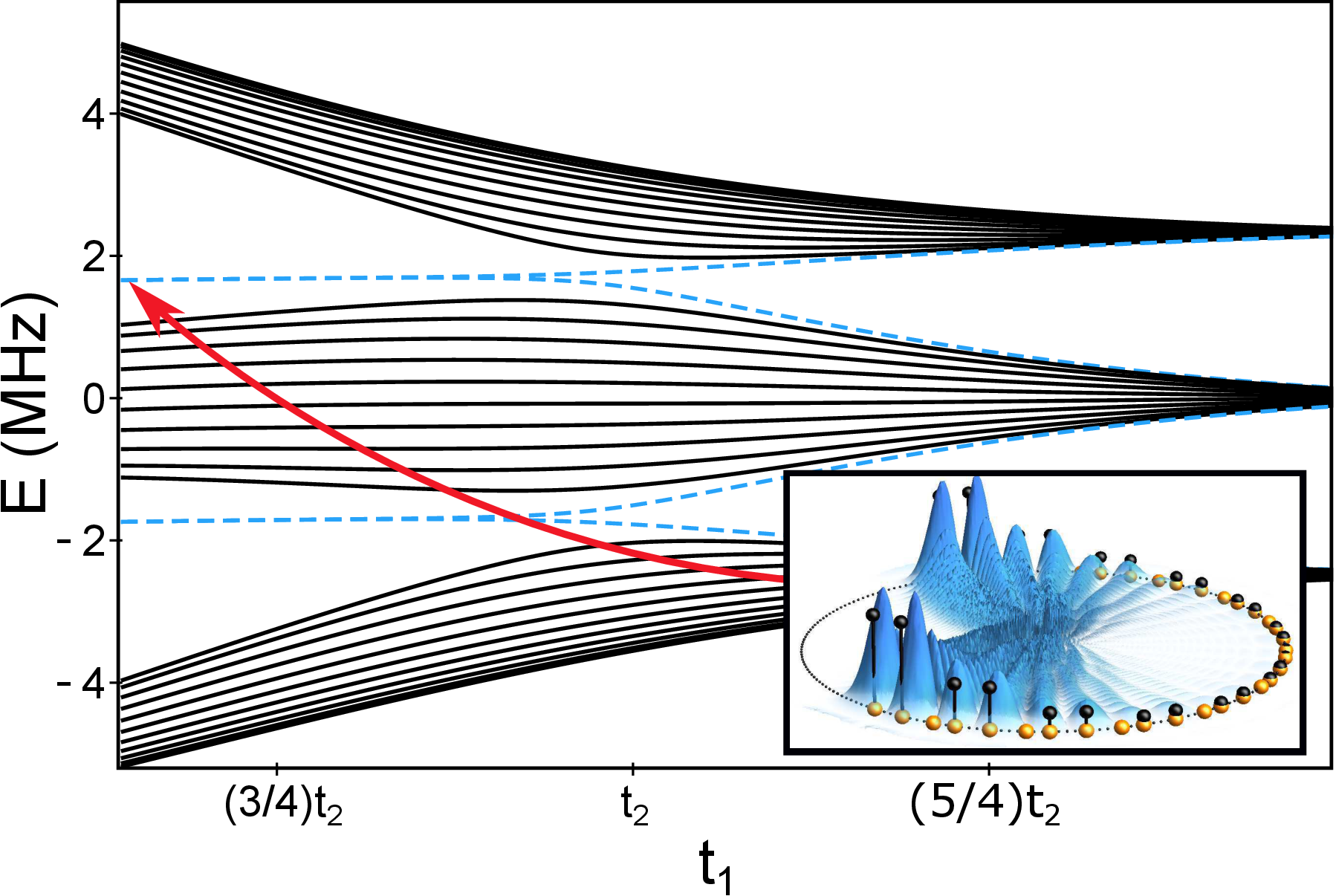}
		\caption{\label{fig:trimers} 
			The energy spectrum of the \ {trimer SSH} Rydberg composite, where $\nu = 60$ and $R = 2\nu^2$. The energies are centered around the on-site potential $E_q$, and are plotted as a function of $t_1$ for fixed $t_2= t_3=2\pi R/45$.   One of the edge states is shown in the inset. 
		}
	\end{figure}

	The second model we consider is a variation on the SSH model where the dimer unit cells are replaced with trimers. 
	This model, shown in Fig.~\ref{fig:intro}(b)i, was very recently explored theoretically \cite{martinezalvarezEdge2019a,anastasiadisBulkedge2022b} as well as experimentally \cite{guoRotation2022,parkCreation2022,kartashovObservation2022}. 
	The sites within each trimer unit cell are coupled by $v_2$ and $v_3$, and the  trimers are coupled via $v_1$. 
	We realize this model in our composite setup by keeping the Rydberg parameters as before, which guarantees nearest neighbor hopping, but now place the scatterers in a different geometry, as seen in Fig. \ref{fig:intro}(b)ii. 
	Applying periodic boundary conditions to the resulting Hamiltonian yields the 
	bulk momentum Hamiltonian
	\be
	H_\text{trimer}(k) = \begin{pmatrix}
		0 & v_2 & v_1e^{-ik}\\ v_2 & 0 & v_3\\
		v_1e^{ik} & v_3 & 0
	\end{pmatrix}.
	\ee
	The topological nature of this model stems from an inversion symmetry when $v_2 = v_3$.
	When $v_1=v_2=v_3$, i.e., in  the absence of trimerization, a phase transition occurs. 
	In the Rydberg composite setting, we ensure the inversion symmetry by setting 
	$t_2=t_3$. By varying $t_1$ to change $v_1$ we sweep the system through the topological phase transition, as can be seen in the energy spectrum shown in Fig.~\ref{fig:trimers}.
	
	The lower band is characterized by {a Zak phase} $\mathcal{Z} = \pi$ when  $v_1>v_2,v_3$ which implies the existence of a pair of topologically protected edge states in the lower band gap. 
	In the Rydberg composite, this energy relationship is satisfied when the distance $t_1$ is smaller than $t_2$ and $t_3$, and we see in Fig.~\ref{fig:trimers} that edge states appear as expected.
	The Zak phase of the middle band is zero for all $v_1$, and since the existence of edge modes in the upper band gap is connected to the sum of the Zak phases of all bands below it,  localized boundary modes must also exist in the upper band gap. 
	One of the edge states is shown in the inset of Fig.~\ref{fig:trimers}. 
	The more complicated case of a non-inversion symmetric lattice, where $v_2 \ne v_3$, can no longer be characterized using an integer Zak phase.
	It has been shown that it supports both topologically-protected and localized, but not protected, edge states \cite{martinezalvarezEdge2019a}.
	Such features can also be simulated and observed in the Rydberg composite by setting $t_2 \ne t_3$.
	\begin{figure*}[t]
		\includegraphics[width=0.94\textwidth]{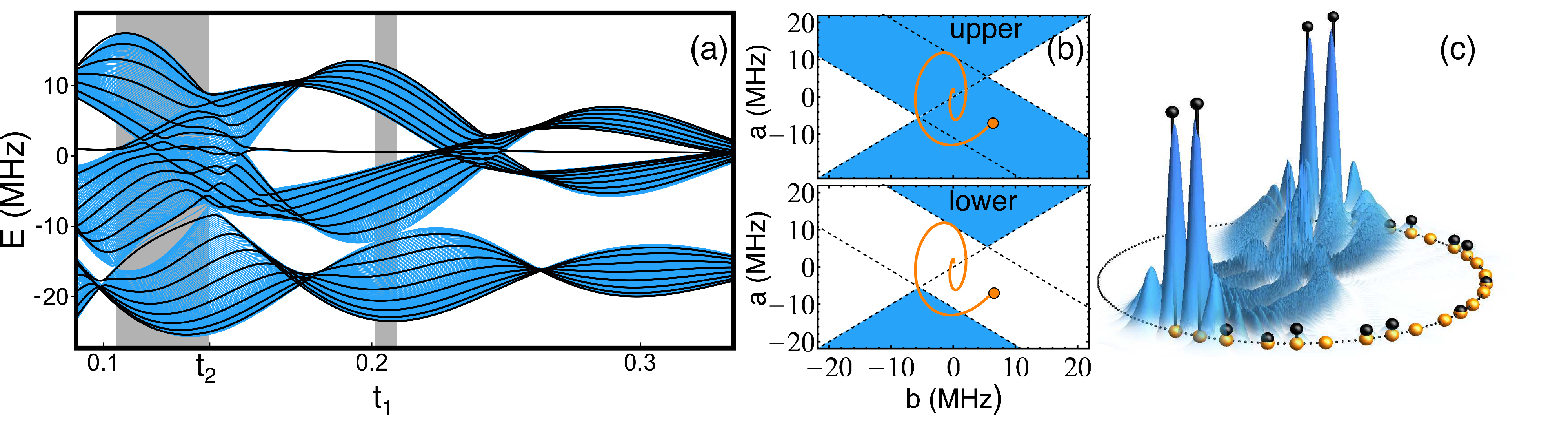}
		\caption{\label{fig:triangles} (a) The energy spectrum of the triangle chain Rydberg composite. 
			The black curves show the Rydberg energies as a function of the arc-length $t_1$ (see Fig.~\ref{fig:intro}(c)ii)  for $\nu = 60$, $R = 1.74\nu^2$, and $M=24$ scatterers.  The energies are centered around the on-site potential $E_q$, and are plotted as a function of $t_1$ for fixed $t_2= 2\pi R/45$.   The blue curves show the spectrum for the same lattice parameters but $M  =300$. Topologically protected edge states are found in the middle of the upper band gap for all $t_1$ values, although they disappear when the bands cross as discussed in the text. One of these edge states, for $t_1= 0.2$, is shown in panel (c). The lower band gap has $\mathcal{Z} = \pi$ only within the gray shaded regions, where degenerate states in the band gap can be seen in the spectrum of the larger lattice. 
			(b) The Zak phase of the upper and lower band gaps. Blue regions have $\mathcal{Z}=\pi$; while white regions have $\mathcal{Z}= 0$. The orange curve shows how this parameter space is traversed as we parameterically change $t_1$, as described in the text.  Note that $u \approx -5.5$MHz for the composite studied here. }
	\end{figure*}
	
	These two examples have demonstrated that the Rydberg composite can realize topologically interesting models, exemplified by the dimer and trimer SSH chains already studied. 
	We now show how to realize a model with richer topology and complexity. 
	We consider again a system with a unit cell of three lattice sites, but now connect each pair of sites within the unit cell with an equal hopping amplitude $u$. 
	Each triangle is coupled to its neighbor by the hopping amplitudes $a$ (magenta), $b$ (green), and $c$ (cyan, dashed), as shown in Fig.~\ref{fig:intro}(c)i; in general these amplitudes can all differ. 
	We realize this Hamiltonian in the Rydberg composite by keeping the same trimer structure, but now utilizing one of the most appealing features of the Rydberg composite: the ease with which we can introduce long-range and oscillatory interactions to the system by changing the overall size of the ring. 
	In general, as the radius of the ring shrinks, the interactions become longer-ranged \cite{anderson}. 
	
	For this example, we select $\nu = 60$ and $R_2 = 1.74\nu^2$, which leads to the interaction curves $V_{qq'}$ shown in Fig.~\ref{fig:intro}(c)iii. 
	These curves are approximately sinuisoidal with a rapidly decreasing amplitude as $|q-q'|$ increases. 
	To see how this choice works to engineer the desired system, consider first the pink curve, which shows the hopping amplitude from site $5$ to its neighbors as a continuous function of the distance $D_5$ away from this site. 
	The pink markers on this curve show the hopping amplitudes connecting site $5$ to the labeled site at the specified positions. 
	The on-site energy $E_5 = V_{55}$ is not visible on this scale. 
	The inversion symmetry of the trimer ensures that $V_{56} = V_{54}$; more distant sites (i.e. $3$ and $7$) are coupled weakly to this site for the chosen value of $t_1$. 
	Now, to make the couplings $V_{46}$ equal to these couplings, we take advantage of the oscillatory interaction. 
	Looking at the black curve, which is the same as the pink curve but relative to site $6$, we can read off the values of $V_{46}$ and $V_{56}$, which are identical as long as the atoms in each unit cell are an arclength $2\pi R_2/45$ apart. 
	In general, larger unit cells with $n$ participating atoms having identical  all-to-all coupling can be engineered by finding parameters such that $V(t_2) = V(2t_2) = \dots V(nt_2)$, where $t_2$ is the arclength between atoms in a cell and the $n+1$th atom is placed further at a $t_1 + nt_2$. 
	It is not clear that such an arrangement can be found for arbitrary $n$, but we verified that the $n=4$ case can be designed, utilizing a smaller ring size where the envelope decays less quickly.

	The band spectrum for this triangle lattice system and $M=24$ atoms, plotted with black curves as a function of $t_1$, is shown  in Fig.~\ref{fig:triangles}(a). 
	Three bands are clearly visible. 
	The upper two repeatedly cross one another.
	Outside of these crossing regions, a pair of zero-energy modes is clearly present in the band gap.  
	Between the lower two bands there are no unambiguous edge states for this small lattice, although inspection of the eigenstates for slightly larger lattice dimensions (not shown here) suggests that the states in between the two bands in the region from $0.1<t_1<t_2$ are indeed edge states. 
	
	To analyze this spectrum, we turn again to the bulk Hamiltonian, which is now given by 
	\be
	H_\text{triangle}^\text{bulk}(k) = \begin{pmatrix}
		2c\cos k & u + b e^{-ik} &u + a e^{-ik}\\ u + b e^{ik}  & 2c\cos k & u+ b e^{-ik}\\
		u+ a e^{ik}&u+ b e^{ik} & 2c\cos k
	\end{pmatrix}.
	\label{eq:HamTriangle}
	\ee
	Note that the coupling $c$ enters the bulk Hamiltonian only on the diagonal and therefore has no impact on the gap closing conditions or the topological phase transitions. 
	The Hamiltonian is inversion-symmetric, and therefore we can characterize its topological features using, again, the Zak phase as a quantized topological invariant. 
	In the bottom (top) panel of  Fig.~\ref{fig:triangles}(c), we show the phase diagram of $\mathcal Z$ as a function of the hopping amplitudes $a$ and $b$ for the lower (upper) band gap. 
	The blue (shaded) color denotes the topological phase ($\mathcal Z = \pi$) and white the trivial  phase ($\mathcal Z = 0$).
	The gray dashed lines show the gap closing conditions, which are sufficient but not necessary for the existence of a topological phase transition.
	These occur when $a=b$ or $b = \pm 2u- a$. 
	We have computed this phase diagram for all $a$ and $b$ values in the plotted range, but as these depend parametrically on the arclength $t_1$, we cannot probe this full parameter space in the Rydberg composite. 
	The orange curve shows the path through this parameter space that we can attain by varying $t_1$  over the range shown in Fig.~\ref{fig:triangles}a, starting at $t_1 = 0.09$ at the marker and ending at $t_1 = 0.33$. 
	Using this curve we can analyze the states in panel (a) with respect to their topological properties.
	For this choice of parameters, the orange curve in the upper band gap does not enter the region with a Zak phase of 0; for this reason there are always edge states visible when the bands are not overlapping. 
	When the bands overlap, these edge states disappear even though the Zak phase remains equal to $\pi$. 
	This surprising observation, contrary to the common belief that a non-trivial Zak phase implies the existence of localized edge states, has been previously observed and discussed in Ref.~\cite{malkiDelocalization2019, malkiAbsence2019}, but it does not seem to be widely known. 
	In contrast, the Zak phase in the bottom gap is zero except for two intervals. 
	These correspond to the regions highlighted in Fig.~\ref{fig:triangles}(a) with gray boxes. 
	Since the spectrum in panel (a) is for only a relatively small number of scatterers, the distinction between band edges and states within the band gap is not clear. 
	Therefore, we artificially extended the system to 300 sites and plotted the resulting spectrum in blue. 
	The edge states in the lower band gap can now clearly be seen, at least in the larger region with $\mathcal{Z} = \pi$ just below $t_1 = t_2$. 
	We have confirmed that these are edge states by adding disorder to the system which preserves the inversion symmetry. 
	Under this disorder, the character of these edge states is preserved. 
	
	The above examples demonstrate that with Rydberg composites one can realize Hamiltonians that exhibit a variety of interesting topological features.
	For the design and interpretation of the Rydberg composite we utilized the useful link between the Rydberg composite and a tight-binding Hamiltonian, which was previously elucidated in the context of Anderson localization \cite{anderson}. 
	While we did not consider disorder in the present work, defects in the positions of the ground state atoms would result in disorder in the hopping terms and on-site energies of the different models studied here, and -- as long as disorder which breaks the chiral or inversion symmetries of our models could be avoided -- the topological protection of the edge states predicted here could be studied. 
	
	Our scheme could be realized using programmable optical tweezers, which have become a very powerful and versatile tool recently.
	Although the parameter regimes used in our examples here are not immediately realizable in current platforms, primarily due to the difficulties in arranging atoms on the $\sim100$ nanometer scale, this is not a fundamental obstacle.
	
	\acknowledgements
	MTE thanks S. Zhang for pointing out Refs. \cite{malkiAbsence2019,malkiDelocalization2019}. 
	C.W.W. acknowledges support from the Max-Planck Gesellschaft via the MPI-PKS Next Step fellowship and is financially supported by the Deutsche Forschungsgemeinschaft (DFG, German Research Foundation) – Project No. 496502542 (WA 5170/1-1).
	AE acknowledges support from the DFG via a Heisenberg fellowship (Grant No EI 872/10-1).

\end{document}